\documentclass{ws-ijmpa}

\begin{document}

\markboth{V. M. Mostepanenko {\it et al.}}
{Why the Screening Effects Do Not Influence the Casimir Force}

%
\catchline{}{}{}{}{}
%

\title{WHY SCREENING EFFECTS DO NOT INFLUENCE THE CASIMIR
FORCE\footnote{This paper is the presentation of two talks given
by V.\ M.\ Mostepanenko and by B.\ Geyer.}}

\author{V. M. MOSTEPANENKO\footnote{On leave from Noncommercial Partnership
``Scientific Instruments'',  Moscow,  Russia}}

\address{Institute for Theoretical Physics, Leipzig University,
D-04009, Leipzig, Germany\\
Vladimir.Mostepanenko@itp.uni-leipzig.de}

\author{R. S. DECCA}

\address{Department of Physics, Indiana University-Purdue
University Indianapolis, IN 46202, USA}

\author{E. FISCHBACH}

\address{Department of Physics, Purdue University, West Lafayette, IN
47907, USA}

\author{ B. GEYER and G. L. KLIMCHITSKAYA\footnote{On leave from
North-West Technical University, St.Petersburg, Russia}}

\address{Institute for Theoretical Physics, Leipzig University,
D-04009, Leipzig, Germany}

\author{D. E. KRAUSE}

\address{Physics Department, Wabash College, Crawfordsville, IN 47933,
USA}

\author{D. L\'{O}PEZ}

\address{Center of Nanoscale Materials,
Argonne National Laboratory, Argonne, IL 60439, USA}

\author{U. MOHIDEEN}

\address{Department of Physics and Astronomy, University of California,
Riverside, CA 92521, USA}

\maketitle

\begin{history}
\received{20 October 2008}
\revised{22 January 2009}
\end{history}

\begin{abstract}
The Lifshitz theory of dispersion forces leads to
thermodynamic and experimental inconsistencies when the role of
drifting charge carriers is included in the model of the dielectric
response. Recently modified reflection coefficients were suggested that
take into account screening effects and diffusion currents.
We demonstrate that this theoretical approach leads to a violation
of the third law of thermodynamics (Nernst's heat theorem) for a wide
class of materials and is excluded by the data from two recent
experiments. The physical reason for its failure is explained by the
violation of thermal equilibrium, which is the
fundamental applicability condition
of the Lifshitz theory, in the presence of drift and diffusion currents.

\end{abstract}

\ccode{PACS numbers: 12.20.-m, 42.50.Ct, 78.20.Ci}

\section{Introduction}

In the last few years the Casimir effect has attracted increasing
attention due to prospective applications in both fundamental physics
and nanotechnology. Sixty years ago H.\ B.\ G,\ Casimir\cite{1} made
his famous discovery that two electrically neutral parallel ideal metal
plates spaced at some separation $a$ in vacuum attract each other.
Casimir explained this effect as due to  the alterations in the spectrum of
zero-point oscillations of the electromagnetic field introduced by
the presence of the plates. In the simplest approach, the Casimir effect
can be described theoretically using quantum field theory with boundary
conditions. The case of real material plates was considered by
Lifshitz,\cite{2} who described material properties by means of a
frequency-dependent dielectric permittivity. The Lifshitz theory was
successfully applied to the interpretation of measurement data in
several experiments measuring the Casimir force.\cite{3}

Further investigations revealed that the Lifshitz theory at nonzero
temperature leads to problems when the relaxation of free charge carriers
is included in the model of the dielectric response. This was shown to lead
to both thermodynamically and
experimental inconsistencies.\cite{4}\cdash\cite{9}
Problems arise in the zero-frequency contribution of the Lifshitz formula
when the free charge carriers are described by the dielectric permittivity
of the Drude model. The Lifshitz theory was found to be thermodynamically
consistent and in agreement with the experimental data if the free electrons
in metals are described by the dielectric permittivity of the plasma
model.\cite{10}\cdash\cite{12} For dielectric and semiconductor materials,
consistency with thermodynamics and experiment is achieved if the dc
conductivity is neglected or charge carriers are described by means of
the plasma model depending on whether the concentration of charge carriers
is below or above the critical value, respectively.\cite{13}

Recently, an alternative approach to the description of free charge carriers
in the Lifshitz theory was suggested\cite{14}\cdash\cite{16} which takes
into account screening effects and diffusion currents. Within this
approach, the macroscopic characteristic of the plate material (the dielectric
permittivity) is supplemented with a microscopic quantity (the density of free
charge carriers). Below we consider the most typical configuration
of two thick parallel
plates (semispaces) and demonstrate that this approach is
thermodynamically and experimentally inconsistent. The reason for this
failure is the violation of thermal equilibrium which is the basic
applicability condition of the Lifshitz theory.

The structure of the paper is as follows. In Sec.~2 we briefly formulate
problems which arise when the Lifshitz theory is applied to materials with
nonzero conductivity. Section~3 contains the formulation of approaches
taking into account the charge screening of free carriers. In Sec.~4 it is
demonstrated that these approaches are in conflict with thermodynamics.
Section~5 is devoted to the comparison of theoretical results obtained
with the inclusion of the screening effects and available experimental data.
In Sec.~6 the reader will find our conclusions and discussion.

\section{Problems of the Lifshitz Theory in Application to
Materials with Nonzero Conductivity}

At nonzero temperature all materials have
a nonzero conductivity. For metals and metallic-type semiconductors
this conductivity can be rather large, and does not go to zero when the
temperature vanishes. For dielectrics and some semiconductors (intrinsic
ones and those with dopant concentration below critical) conductivity is
much smaller than for metals, and goes to zero together with temperature.
In the Lifshitz theory,
the free energy per unit area in the configuration of two
semispaces described by the dielectric permittivity
$\varepsilon(\omega)$ is given by
\begin{eqnarray}
&&
{\cal F}(a,T)=\frac{k_BT}{2\pi}\sum_{l=0}^{\infty}
{\vphantom{\sum}}^{\prime}\int_{0}^{\infty}\!\!k_{\bot}dk_{\bot}
\left\{\ln\left[1-r_{\rm TM}^2({\rm i}\xi_l,k_{\bot}){\rm e}^{-2aq_l}\right]
\right.
\label{eq1} \\
&&~~~~~~~~~~~~~~~~~~~~~~~\left.
+\ln\left[1-r_{\rm TE}^2({\rm i}\xi_l,k_{\bot}){\rm e}^{-2aq_l}\right]\right\}.
\nonumber
\end{eqnarray}
\noindent
Here, $a$ is the separation distance between the semispaces, $k_{\bot}$
is the magnitude of the wave vector in the plane of boundary plates,
the primed sum adds a multiple 1/2 to the term with $l=0$,
$\xi_l=2\pi k_BTl/\hbar$ with $l=0,\,1,\,2,\,\ldots$ are the Matsubara
frequencies, and $k_B$ is the Boltzmann constant. Equation (\ref{eq1})
is derived under the condition that the plates (semispaces)
are at a temperature $T$ in thermal equilibrium with
the environment. The reflection coefficients
for two independent polarizations of the electromagnetic field
coincide with the
Fresnel ones calculated along the imaginary frequency axis,
\begin{equation}
r_{\rm TM}({\rm i}\xi_l,k_{\bot})=\frac{\varepsilon_lq_l-
k_l}{\varepsilon_lq_l+k_l}, \qquad
r_{\rm TE}({\rm i}\xi_l,k_{\bot})=\frac{q_l-k_l}{q_l+k_l},
\label{eq2}
\end{equation}
\noindent
where
\begin{equation}
q_l^2=k_{\bot}^2+\frac{\xi_l^2}{c^2},\qquad
k_l^2=k_{\bot}^2+\varepsilon_l\frac{\xi_l^2}{c^2},\qquad
\varepsilon_l\equiv\varepsilon({\rm i}\xi_l).
\label{eq3}
\end{equation}

Originally, Eqs.~(\ref{eq1})--(\ref{eq3}) were mostly applied to dielectrics
with the  dc conductivity neglected. For such materials the dielectric
permittivity can be written in the form\cite{17}
\begin{equation}
\varepsilon({\rm i}\xi)=1+\sum_{j=1}^{K}
\frac{f_j}{\omega_j^2+\xi^2+\gamma_j\xi},
\label{eq4}
\end{equation}
\noindent
where $\omega_j\neq 0$ are the oscillator frequencies, $f_j$ are the
oscillator strengths and $\gamma_j$ are the relaxation parameters.
This equation describes the dielectric response of core electrons.
The free energy (\ref{eq1}) with the dielectric permittivity (\ref{eq4})
is in perfect agreement with thermodynamics. Specifically, the Casimir
entropy vanishes with temperature.\cite{13}

Problems with thermodynamics arise when the conductivity $\sigma$ of
the plate
material is taken into account. For metals and metallic-type semiconductors
the dielectric permittivity can be modelled by means of the Drude model,
\begin{equation}
\tilde\varepsilon({\rm i}\xi)=\varepsilon({\rm i}\xi)+
\frac{4\pi\sigma({\rm i}\xi)}{\xi}=\varepsilon({\rm i}\xi)+
\frac{\omega_p^2}{\xi(\xi+\gamma)},
\label{eq5}
\end{equation}
\noindent
where $\omega_p$ is the plasma frequency and $\gamma$ is the relaxation
parameter of free electrons. Simple expressions for the quantities
entering Eq.~(\ref{eq5}) are:\cite{18}
\begin{equation}
\sigma({\rm i}\xi)=\frac{\sigma(0)}{1+\frac{\xi}{\gamma}}, \qquad
\omega_p^2=\frac{4\pi e^2n}{m}, \qquad
\sigma(0)=\mu|e|n,
\label{eq6}
\end{equation}
\noindent
where $\sigma(0)$ is the dc conductivity, $e$ and $m$ are the charge and
effective mass
of the electron, $n$ is the density of charge carriers and $\mu$ is their
mobility. The substitution of the dielectric permittivity (\ref{eq5})
into Eqs.~(\ref{eq1})--(\ref{eq3}) results in a negative Casimir
entropy at $T=0\,$K depending on $a$ for metals with perfect crystal
lattices,\cite{4} which is
 in violation of the third law of thermodynamics
(the Nernst heat theorem).
Recently it was claimed\cite{18a} that Ref.~\refcite{4} is wrong
because it uses the theory of the normal skin effect for metals with
perfect crystal lattices at $T\to 0$, while in this situation one
must use the theory of the anomalous skin effect. This objection is,
however, incorrect. With the decrease of $T$ the application
region of the normal skin effect does become narrower and
the application region of the anomalous skin effect widens.
However, at any $T>0$, there exists a frequency region near zero
frequency, where the normal skin effect is applicable.
Keeping in mind that the violation of the Nernst heat theorem
originates entirely from the zero-frequency term of the Lifshitz
formula, one concludes that the permittivity of the normal skin
effect is appropriate for the evaluation of this term at low $T$
when considering the thermodynamic consistency of the Lifshitz
theory. There is a
suggestion\cite{19,20} to satisfy the Nernst heat theorem by the inclusion
of impurities, but this does not solve the problem\cite{21,22}
because according to quantum statistical physics for perfect crystal
lattices the Casimir entropy at zero temperature must be equal to zero.

Another problem is the experimental inconsistency of the Lifshitz theory
combined with the dielectric permittivity (\ref{eq5}). This was demonstrated
in a series of three successive experiments on the dynamic determination
of the Casimir pressure between two gold plates.\cite{6}\cdash\cite{8,23}
For example, in Fig.~1 we demonstrate the comparison between the
measurement data of the most precise third experiment\cite{7,8} and
theoretical Casimir pressures computed using the Drude model (\ref{eq5})
and the generalized plasma-like model of Refs.~\refcite{10}--\refcite{12}
[dielectric permittivity (\ref{eq5}) with $\gamma=0$].
In Fig.~1(a) the Casimir pressures computed using the Drude model
approach and the generalized plasma-like
model are shown  as dark-gray and light-gray bands, respectively.
The experimental data are shown as crosses. The widths of the bands and the
sizes of the arms of the crosses are determined at a 95\% confidence
level. In Fig.~1(b), the differences of the theoretical Casimir pressures
computed using Eq.~(\ref{eq5}), $P_{D}^{\rm theor}(a)$, and the mean
experimental Casimir pressures, $\bar{P}^{\rm expt}(a)$, are shown as dots
labeled 2. The differences of the theoretical pressures computed using the
generalized plasma-like permittivity [Eq.~(\ref{eq5}) with $\gamma=0$],
$P_{gp}^{\rm theor}(a)$, and $\bar{P}^{\rm expt}(a)$ are shown as dots
labeled 1. The solid and dashed lines indicate the boundaries
of 95\%  and 99.9\% confidence intervals, respectively.
As can be seen in Fig.~1(a,b), the generalized plasma-like model is consistent
with data, whereas the Drude permittivity taking into account the relaxation
properties connected with a drift current of conduction electrons is
experimentally excluded.
\begin{figure}[pt]
\vspace*{-9.5cm}
\centerline{\psfig{file=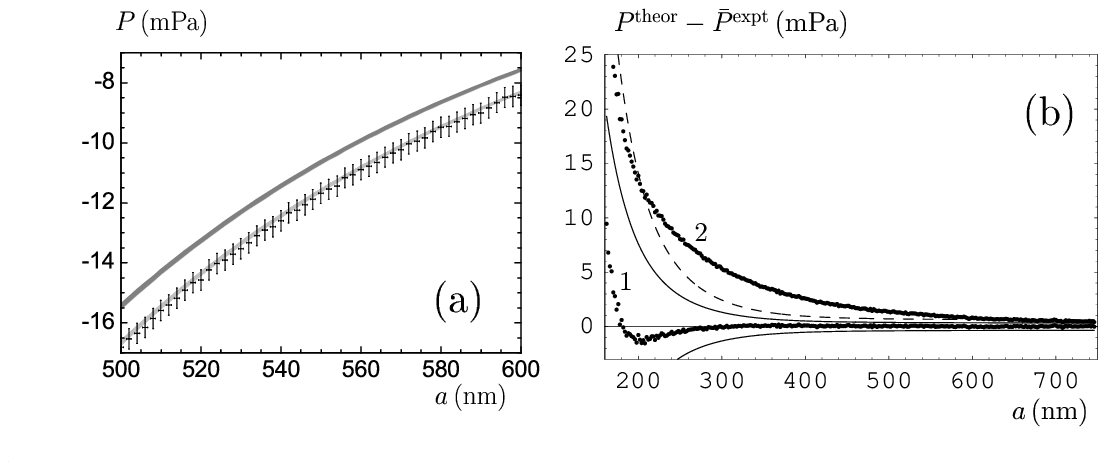,width=22cm}}
\vspace*{-17.4cm}
\caption{(a) The crosses show the measured mean Casimir pressures together
with the absolute errors
in the separation and pressure as a function of the separation.
The theoretical Casimir pressures computed using
the generalized plasma-like model and the optical data extrapolated by
the Drude model are shown by the light-gray and dark-gray bands,
respectively.
(b) The differences of the
theoretical and the  mean experimental
Casimir pressures between two Au plates
versus separation are shown as dots.
The theoretical results are calculated using the
Lifshitz theory at room temparature using the generalized plasma-like
model (the dots labeled 1) and the Drude model approach (the dots labeled 2).
The solid and dashed lines indicate the boundaries
of  95\% and 99.9\% confidence intervals, respectively.}
\end{figure}

\begin{figure}[pb]
\vspace*{-4.7cm}
\centerline{\hspace*{-15mm}\psfig{file=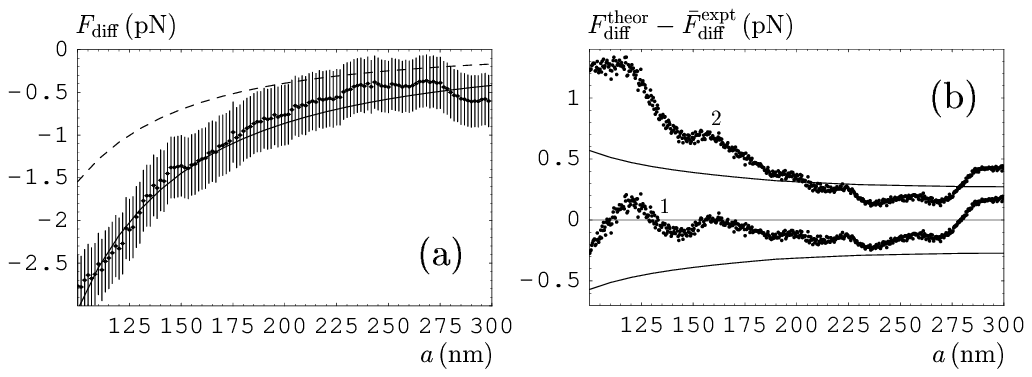,width=24cm}}
\vspace*{-25.3cm}
\caption{(a) The measured force differences are shown as crosses
versus separation $a$. The solid and dashed lines present
 the theoretical results calculated with  the neglected and included
dc conductivity of high-resistivity Si in the dark phase,
respectively.
(b) Theoretical minus mean experimental differences of the
Casimir force for the Lifshitz theory with neglected (label 1) and
included (label 2) dc conductivity of high-resistivity Si in the dark phase.
The solid lines indicate 95\% confidence intervals.
}
\end{figure}
Now we consider dielectric plates, whose conductivity vanishes with $T$, and
take into account the small dc conductivity using
the first equality in Eq.~(\ref{eq5}) in the
calculation of the Casimir free  energy and entropy
[the second  equality in Eq.~(\ref{eq5}) is related only to metals].
In this case the
Casimir entropy at $T=0$ takes a positive value
depending on $a$,\cite{5,13,24,25} i.e.,
the Nernst heat theorem is violated. Theoretical results computed with
inclusion of the dc conductivity of the dielectric are in disagreement with the
experimental results on the measurement of the Casimir force between a gold
sphere and a silicon plate illuminated with laser pulses.\cite{26,27}
As an illustration, in Fig.~2(a) the mean measured differences of the
Casimir forces in the presence and in the absence of light on the
plate (the absorbed
power is equal to 4.7\,mW),
$\bar{F}_{\rm diff}^{\rm expt}$, are shown as crosses (the arm
sizes are determined at a 95\% confidence level). The theoretical
differences, ${F}_{\rm diff}^{\rm theor}$, shown by the solid line, are
computed by using the dielectric permittivity of the generalized plasma-like
model in the presence of light and neglecting the dc conductivity in the
absence of light on the plate. The theoretical differences shown by
the dashed line are computed by using the dielectric permittivity (\ref{eq5})
with the respective values of $\sigma$ in the presence, and in the absence, of
light, i.e., with included dc conductivity of the silicon plate also in the
dark phase. As is seen in Fig.~2(a), the theory taking into account the
conductivity of Si in the dark phase using Eq.~(\ref{eq5})
is experimentally excluded, whereas
the theory neglecting the dc conductivity of
dielectric silicon is consistent with the data.

The comparison of experiment with different theories can also be done in
another manner. In Fig.~2(b) we plot as dots the theoretical minus mean
experimental differences of the Casimir forces,
${F}_{\rm diff}^{\rm theor}-\bar{F}_{\rm diff}^{\rm expt}$,
in the presence and in the absence of light. For the dots labeled 2
the values of ${F}_{\rm diff}^{\rm theor}$ are computed with the dc
conductivity of Si included in the dark phase, and for dots labeled 1 the dc
conductivity is neglected. The solid lines indicate the boundaries of
95\% confidence intervals. As is seen in Fig.~2(b), the theory including the
dc conductivity in the dark phase is excluded by the data over a wide
separation region, whereas the
theory neglecting the dc conductivity is experimentally consistent.

The physical reasons why the use of the dielectric permittivity (\ref{eq5})
in the Lifshitz theory leads to problems with thermodynamics and experiment
are discussed in Ref.~\refcite{28}. The point is that this permittivity
describes the influence of the drift current of conduction electrons,
 an irreversible process. It is accompanied by
Joule losses and the heating of the Casimir plates. To keep the temperature
constant, one should allow the existence of an unidirectional flux of
heat from the plates to
the heat reservoir.\cite{29} This is a situation out of
thermal equilibrium.\cite{30} Thus, the dielectric permittivity (\ref{eq5})
describes processes  violating thermal equilibrium which is the basic
applicability condition of the Lifshitz theory. It is thus not surprising
that in combination with the permittivity (\ref{eq5}),  the Lifshitz theory
becomes thermodynamicly and experimentally inconsistent.

\section{Attempts to Generalize the Lifshitz Theory Through the
Inclusion of Screening Effects}

As discussed in Sec.~2, to avoid problems in the application of the Lifshitz
theory to real materials one should neglect the dc conductivity of
dielectrics and describe charge carriers in metals using the generalized
plasma-like dielectric permittivity. Another approach to the resolution
of problems arising in the Lifshitz theory when it is applied to
materials with nonzero conductivity is by including the effect of
Coulomb screening by mobile charges.
As discussed above, at nonzero temperature all materials contain charge
carriers which can move through the medium.
In the presence of mobile charge carriers the standard Coulomb
potential of some external charge is replaced with the so-called
{\it screened} Coulomb potential having a Yukawa-type dependence on
separation.\cite{31} The effect of screening leads to penetration of
a static electric field into materials with nonzero conductivity
to a depth of the screening length which is very small for good metals
(of order the interatomic distance), but can be much larger
for semiconductors.

The response of charge carriers to a spatially variable, sinusoidally
varying, external electric field
taking into account scattering can be described
by Boltzmann transport equation. This approach was recently used\cite{15}
to describe the interaction of the fluctuating electromagnetic field with
mobile charge carriers in a semiconductor plate and to find the modified
reflection coefficients for the TM and TE modes. The resulting coefficients
were substituted into the standard Lifshitz formula (\ref{eq1}).
Here, we consider the applicability of this approach to different materials
leaving the problems of its consistency with thermodynamics and
experiment for the following sections.

The Boltzmann transport equation is used for the description of irreversible
nonequilibrium processes, and is not symmetric with respect to time
reversal. As a result, it describes
processes which lead to an increase of entropy.\cite{32}
In the case under consideration this equation includes both the drift
current of charge carriers $\mbox{\boldmath$j$}$ and the diffusion current
$eD\nabla n$, where $D$ is the diffusion coefficient. As a result,
the modified TM reflection coefficient takes the form\cite{15}
\begin{equation}
r_{\rm TM}^{\rm mod}({\rm i}\xi_l,k_{\bot})=\frac{\tilde\varepsilon_lq_l-
k_l-k_{\bot}^2\eta_l^{-1}\varepsilon_l^{-1}(\tilde\varepsilon_l-
\varepsilon_l)}{\tilde\varepsilon_lq_l+
k_l+k_{\bot}^2\eta_l^{-1}\varepsilon_l^{-1}(\tilde\varepsilon_l-
\varepsilon_l)}.
\label{eq7}
\end{equation}
\noindent
Here, $\varepsilon_l$ and $q_l$ are defined in Eq.~(\ref{eq3}),
$\tilde\varepsilon_l\equiv\tilde\varepsilon({\rm i}\xi_l)$ is
defined in Eq.~(\ref{eq5}), $k_l$ is given by Eq.~(\ref{eq3}) where
$\varepsilon_l$ is replaced with $\tilde\varepsilon_l$ and
\begin{equation}
\eta_l\equiv\eta({\rm i}\xi_l)=\left[k_{\bot}^2+\kappa^2\,
\frac{\varepsilon_0}{\varepsilon_l}\,
\frac{\tilde\varepsilon_l}{\tilde\varepsilon_l-\varepsilon_l}\right]^{1/2},
\qquad \varepsilon_0\equiv\varepsilon(0).
\label{eq8}
\end{equation}
\noindent
 The screening length $1/\kappa$ in Eq.~(\ref{eq8}) is different for different
types of statistics of charge carriers (see below). The modified TE
reflection coefficient, $r_{\rm TE}^{\rm mod}({\rm i}\xi_l,k_{\bot})$,
is given by the standard expressions (\ref{eq2}) and (\ref{eq3}), where
the permittivity $\varepsilon_l$ is replaced by $\tilde\varepsilon_l$.

It has been shown\cite{33} that the problems of the standard Lifshitz theory
discussed in Sec.~2 are mathematically connected with the discontinuity
of the zero-frequency term of the Lifshitz formula with respect to
frequency and temperature at the point (0,0). The modified reflection
coefficients at zero frequency take the form
\begin{equation}
r_{\rm TM}^{\rm mod}(0,k_{\bot})=\frac{\varepsilon_0
\sqrt{k_{\bot}^2+\kappa^2}-k_{\bot}}{\varepsilon_0
\sqrt{k_{\bot}^2+\kappa^2}+k_{\bot}}, \qquad
r_{\rm TE}^{\rm mod}(0,k_{\bot})=0.
\label{eq9}
\end{equation}
\noindent
Note that the TM coefficient in (\ref{eq9}) was first obtained by
Pitaevskii\cite{14} within the theoretical approach taking into account the
penetration of the static component of the fluctuating field in the material
of a wall in the atom-wall (Casimir-Polder) interaction.
According to (\ref{eq9}),
the modified TE reflection coefficient at zero frequency takes the same value
as in the Drude model approach.
Mathematically, this suggests, that for metals the theoretical
approach taking into account screening effects should face the same
difficulties as the Drude model approach (see Sec.~2).

The reflection coefficients $r_{\rm TM,TE}^{\rm mod}(0,k_{\bot})$
can be obtained\cite{14,28} as the standard Fresnel reflection coefficients
for an uniaxial crystal,
\begin{equation}
r_{\rm TM}^{u}({\rm i}\xi_l,k_{\bot})=
\frac{\sqrt{\varepsilon_{xl}\varepsilon_{zl}}q_l-
k_{zl}}{\sqrt{\varepsilon_{xl}\varepsilon_{zl}}q_l+k_{zl}},
\qquad
r_{\rm TE}^{u}({\rm i}\xi_l,k_{\bot})=
\frac{q_l-k_{xl}}{q_l+k_{xl}},
\label{eq10}
\end{equation}
\noindent
where
\begin{equation}
k_{xl}^2=k_{\bot}^2+\varepsilon_{xl}\frac{\xi_l^2}{c^2}, \quad
k_{zl}^2=k_{\bot}^2+\varepsilon_{zl}\frac{\xi_l^2}{c^2}, \quad
\varepsilon_{xl}=\varepsilon_x({\rm i}\xi_l), \quad
\varepsilon_{zl}=\varepsilon_z({\rm i}\xi_l),
\label{eq11}
\end{equation}
\noindent
if one introduces the dielectric permittivity depending on the wave vector
$k_{\bot}$,
\begin{equation}
\varepsilon_{x0}\equiv\varepsilon_{x}(0)=\varepsilon_0,\qquad
\varepsilon_{z0}\equiv\varepsilon_{z}(0)=\varepsilon_0\left(1+
\frac{\kappa^2}{k_{\bot}^2}\right).
\label{eq12}
\end{equation}
\noindent
This means that the standard Lifshitz formula (at least its zero-frequency
term) is applied to spatially dispersive materials. Such an application
is controversial and has been debated in the literature.\cite{34}
The modified  reflection coefficients
$r_{\rm TM,TE}^{\rm mod}({\rm i}\xi_l,k_{\bot})$ at any frequency are also
claimed to be obtainable in terms of spatially nonlocal dielectric
functions.\cite{15} It is easily seen, however, that there are no
such dielectric functions $\varepsilon_{xl}$ and $\varepsilon_{zl}$ which
transform the reflection coefficients (\ref{eq10}) for an uniaxial
crystal into
$r_{\rm TM,TE}^{\rm mod}({\rm i}\xi_l,k_{\bot})$. To prove this, we first
require that $r_{\rm TE}^{u}({\rm i}\xi_l,k_{\bot})$ be equal to
$r_{\rm TE}^{\rm mod}({\rm i}\xi_l,k_{\bot})$. For this purpose one must
put $\varepsilon_{xl}\equiv\tilde\varepsilon_{l}$. Then, keeping in mind
that $\tilde\varepsilon_{l}$ goes to infinity when $\xi\to 0$, one finds
that in this limiting case $r_{\rm TM}^{u}({\rm i}\xi_l,k_{\bot})$ in
Eq.~(\ref{eq10}) goes to unity independent of the functional form of
$\varepsilon_{zl}$. Thus, there is no such function $\varepsilon_{zl}$
that would lead to the required equality
$r_{\rm TM}^{u}(0,k_{\bot})=r_{\rm TM}^{\rm mod}(0,k_{\bot})$,
as defined in Eq.~(\ref{eq9}).

In Ref.~\refcite{34a} the modified reflection coefficients
$r_{\rm TM,TE}^{\rm mod}({\rm i}\xi_l,k_{\bot})$
at any frequency are expressed in terms of two dielectric
functions $\varepsilon_x({\rm i}\xi_l,k_{\bot})$ and
$\varepsilon_z({\rm i}\xi_l,k_{\bot})$ defined in the random phase
approximation.\cite{34b} The obtained expression for
$\varepsilon_z({\rm i}\xi_l,k_{\bot})$ is, however, incorrect. It
does not transform into the dielectric permittivity of dielectric
material $\varepsilon({\rm i}\xi_l)$ defined in Eq.~(\ref{eq4}) in
the limiting case $n\to 0$. This is caused by several mistakes in
sign made in Ref.~\refcite{34a} and by the disagreement of phase
multiples in the TM reflection coefficients used by the authors
of Ref.~\refcite{34a} and in the method of random
phase approximation. After correcting these errors one obtains
\begin{eqnarray}
&&
\varepsilon_z({\rm i}\xi_l,k_{\bot})=k_{\bot}\left[
\frac{k_l\varepsilon_l+k_{\bot}^2\eta_l^{-1}
(\tilde\varepsilon_l-\varepsilon_l)}{\varepsilon_l\tilde\varepsilon_l}+
k_{\bot}-q_l-\frac{\xi_l^2}{c^2}\left(\frac{1}{k_l}-
\frac{1}{q_l}\right)
\right.
\nonumber \\
&&
~~~~~~~~~~~~~~~~~~\left.
+k_{\bot}\frac{\xi_l^2}{c^2}\left(
\frac{1}{k_{\bot}k_l+k_l^2}-\frac{1}{k_{\bot}q_l+q_l^2}\right)
\right]^{-1}\!\!\!.
\label{eq12a}
\end{eqnarray}
\noindent
Here, $k_l$ is defined by Eq.~(\ref{eq3}) where $\varepsilon_l$
is replaced with $\varepsilon_x({\rm i}\xi_l,k_{\bot})$.
The latter coincides with $\tilde\varepsilon({\rm i}\xi_l)$
from Eq.~(\ref{eq5}), i.e., it does not depend on $k_{\bot}$.
It is easily seen that in the limiting case $n\to 0$ the
permittivity (\ref{eq12a}) transforms into
$\varepsilon_l=\varepsilon({\rm i}\xi_l)$.

We emphasize that the coefficient
$r_{\rm TM}^{\rm mod}({\rm i}\xi_l,k_{\bot})$ obtained in
the random phase approximation
 does not coincide with the coefficient
$r_{\rm TM}^{u}({\rm i}\xi_l,k_{\bot})$ for an uniaxial crystal,
as given by Eq.~(\ref{eq10}). It is notable also that the
representation for the reflection coefficient
$r_{\rm TM}^{\rm mod}({\rm i}\xi_l,k_{\bot})$  by means of the
dielectric permittivity $\varepsilon_z({\rm i}\xi_l,k_{\bot})$
at both zero and nonzero frequency is a phenomenological one.
The point is that in the presence of a gap between the plates
the translational invariance in $z$-direction, i.e., perpendicular
to the plates, is violated and $\varepsilon_z({\rm i}\xi_l,k_{\bot})$
is an ill-defined quantity.

Phenomenological reflection coefficients at any frequency, depending on two
spatially nonlocal dielectric permittivities $\varepsilon_{xl}$,
$\varepsilon_{zl}$ and the screening length $1/\kappa$, were also
suggested without refer to Boltzmann transport equation.\cite{16}
At zero frequency they take the same values as in
Eq.~(\ref{eq9}). This approach is a crude approximation.
Given the absence of
translational invariance along the $z$-axis mentioned above,
 it is impossible to define the dielectric permittivity
$\varepsilon_{zl}$ depending on the frequency and the wave vector.
Additionally, specular  reflection of charge carriers on the boundary
planes was assumed.\cite{16} However, for
spatially dispersive materials the scattering of carriers is neither specular
nor diffuse.\cite{35} Because of this, it was concluded\cite{36} that
this approach does not contain self-consistent checks of its accuracy,
and hence could only be justified based on  agreement with
the experimental data and
fundamental physical principles. In the next two sections we demonstrate,
however, that theoretical approaches taking the screening effects
into account are in disagreement with thermodynamics and available
experimental data.

In the end of this section we discuss the claimed application region of the
modified reflection coefficients taking into account screening effects.
This is a familiar subject when describing conducting materials in an
external electromagnetic field (we recall that Ref.~\refcite{15} considers
the fluctuating field in a similar way as an external one).
In Ref.~\refcite{15}, the Boltzmann transport equation is
applied in the nondegenerate continuum limit for sufficiently low
density of charge carriers $n$ (intrinsic semiconductors). In this case
the screening length is given by a specific Debye-H\"{u}ckel
expression,
\begin{equation}
\frac{1}{\kappa}=\frac{1}{\kappa_{\rm DH}}=
\sqrt{\frac{\varepsilon_0k_BT}{4\pi e^2n}}.
\label{eq13}
\end{equation}
\noindent
In fact this expression is applicable to charge carriers obeying
Maxwell-Boltzmann statistics. It is obtained from the general representation
for the screening length,\cite{31}
\begin{equation}
\frac{1}{\kappa}=
\sqrt{\frac{\varepsilon_0 D}{4\pi \sigma(0)}},
\label{eq14}
\end{equation}
\noindent
if one uses $\sigma(0)$ from Eq.~(\ref{eq6}) and Einstein's
relation\cite{18,31} $D/\mu=k_BT/|e|$ valid in the case of
Maxwell-Boltzmann statistics.

However, the application region of the modified reflection coefficients
$r_{\rm TM,TE}^{\rm mod}$ with the Debye-H\"{u}ckel screening length
(\ref{eq13}) is not restricted to only intrinsic semiconductors, but
they are applicable to all materials where $n$ is not too large
so that charge carriers are described by Maxwell-Boltzmann statistics.
This means that it is legitimate to apply the coefficients
$r_{\rm TM,TE}^{\rm mod}$ to doped semiconductors with dopant concentration
below critical, and to solids with ionic conductivity, etc.

Reference \refcite{15} claims that its approach ``does not apply
to metals, where the electron density is sufficiently large and the
electron gas is degenerate''. The Boltzmann transport equation, however,
is equally applicable to classical and quantum systems. The only difference
one should take into account is the type of statistics. In metals and
metallic-type semiconductors  charge carriers obey the quantum
Fermi-Dirac statistics. Substituting Einstein's relation valid in the case
of Fermi-Dirac statistics,\cite{18,31} $D/\mu=2E_F/(3|e|)$, where
$E_F=\hbar\omega_p$ is the Fermi energy, into Eq.~(\ref{eq14}), one
arrives at the Thomas-Fermi screening length,\cite{31}
\begin{equation}
\frac{1}{\kappa}=\frac{1}{\kappa_{\rm TF}}=
\sqrt{\frac{\varepsilon_0E_F}{6\pi e^2n}}.
\label{eq15}
\end{equation}
\noindent
With this definition of the parameter $\kappa$, it is justified to apply
the modified reflection coefficients $r_{\rm TM,TE}^{\rm mod}$ to metals.
Thus, with the proper definition of the screening length, the suggested
generalization of the Lifshitz theory, if at all meaningful,
should be applicable  to any material. The problem
remains whether the screening effects are relevant to the Casimir force.
This is discussed in the next sections.

\section{Inclusion of Screening Effects and Thermodynamics}

Here we perform the thermodynamic test of the Lifshitz formula (\ref{eq1})
combined with the modified reflection coefficients  $r_{\rm TM}^{\rm mod}$
from Eqs.~(\ref{eq7}), (\ref{eq8}) and   $r_{\rm TE}^{\rm mod}$.
For this purpose the Casimir entropy at zero temperature will be calculated
analytically. It is convenient to introduce the dimensionless variables
$y=2aq_l$ and $\zeta_l=\xi_l/\omega_c\equiv 2a\xi_l/c$. Then the
modified Casimir free energy takes the form
\begin{eqnarray}
&&
{\cal F}^{\rm mod}(a,T)=\frac{k_BT}{8\pi a^2}\sum_{l=0}^{\infty}
{\vphantom{\sum}}^{\prime}\int_{\zeta_l}^{\infty}\!\!y\,dy
\left\{\ln\left[1-
r_{\rm TM}^{\rm mod}{\vphantom{r^{\rm d}}}^2({\rm i}\zeta_l,y)
{\rm e}^{-y}\right]
\right.
\label{eq16} \\
&&~~~~~~~~~~~~~~~~~~~~~~~\left.
+\ln\left[1-r_{\rm TE}^{\rm mod}{\vphantom{r^{\rm d}}}^2({\rm i}\xi_l,y)
{\rm e}^{-y}\right]\right\}.
\nonumber
\end{eqnarray}
\noindent
In terms of dimensionless variables the reflection coefficient
(\ref{eq7}) can be rearranged as
\begin{equation}
{r}_{\rm TM}^{\rm mod}({\rm i}\zeta_l,y)=\frac{\tilde\varepsilon_l y
-\bigl[y^2+(\tilde\varepsilon_l-1)\zeta_l^2\bigr]^{1/2}-
(y^2-\zeta_l^2)(\tilde\varepsilon_l-
\varepsilon_l)\tilde\eta_l^{-1}
\varepsilon_l^{-1}}{\tilde\varepsilon_l y
+\bigl[y^2+(\tilde\varepsilon_l-1)\zeta_l^2\bigr]^{1/2}+
(y^2-\zeta_l^2)(\tilde\varepsilon_l-
\varepsilon_l)\tilde\eta_l^{-1}
\varepsilon_l^{-1}},
\label{eq17}
\end{equation}
\noindent
where
\begin{equation}
\tilde\eta_l=2a\eta_l=\left[y^2-\zeta_l^2+\kappa_a^2\frac{\varepsilon_0
\tilde\varepsilon_l}{\varepsilon_l
(\tilde\varepsilon_l-\varepsilon_l)}
\right]^{1/2}, \qquad
\kappa_a\equiv 2a\kappa.
\label{eq18}
\end{equation}
\noindent
Note that all dielectric permittivities here are functions of
${\rm i}\omega_c\zeta_l$.
Below we do not use the explicit expression for the reflection
coefficient $\tilde{r}_{\rm TE}({\rm i}\zeta_l,y)$ because it coincides
with the standard one, as defined in the Drude model approach, which was
considered in detail in Ref.~\refcite{4}.

Let us determine the behavior of the Casimir free energy (\ref{eq16}) at
low temperatures.
We begin with the case of metals where $\kappa=\kappa_{\rm TF}$.
For all metals the screening length is very
small. As a result, at any reasonable separation distance between the
plates, the dimensionless parameter $\kappa_a$ defined in (\ref{eq18})
is very large and the inverse quantity $\beta_a\equiv 1/\kappa_a\ll 1$
can be used as a small parameter.
Expanding the reflection coefficient (\ref{eq17}) up to the first power
of the parameter $\beta_a$ one obtains
\begin{eqnarray}
&&
{r}_{\rm TM}^{\rm mod}({\rm i}\zeta_l,y)=\tilde{r}_{\rm TM}({\rm i}\zeta_l,y)-
2\beta_a\,Z_l+O(\beta_a^2),
\label{eq19} \\
&&
Z_l\equiv\sqrt{\frac{\tilde\varepsilon_l
(\tilde\varepsilon_l-\varepsilon_l)^3}{\varepsilon_0\varepsilon_l}}
\,\frac{y(y^2-\zeta_l^2)}{[\tilde\varepsilon_l y+\sqrt{y^2+
(\tilde\varepsilon_l -1)\zeta_l^2}]^2},
\nonumber
\end{eqnarray}
\noindent
where $\tilde{r}_{\rm TM}({\rm i}\zeta_l,y)$ is the standard TM reflection
coefficient calculated with the dielectric permittivity
$\tilde\varepsilon({\rm i}\omega_c\zeta_l)$. [It is given by Eq.~(\ref{eq17})
with the third terms in both the numerator and the denominator omitted.]
{}From Eq.~(\ref{eq19}) one arrives at
\begin{equation}
\ln\left[1-{r}_{\rm TM}^{\rm mod}{\vphantom{r^{\rm d}}}^2({\rm i}\zeta_l,y)
\,{\rm e}^{-y}\right]=
\ln\left[1-\tilde{r}_{\rm TM}^2({\rm i}\zeta_l,y)\,{\rm e}^{-y}\right]+
4\beta_a\frac{\tilde{r}_{\rm TM}({\rm i}\zeta_l,y)\,Z_l}{{\rm e}^{y}-
\tilde{r}_{\rm TM}^2({\rm i}\zeta_l,y)}+O(\beta_a^2).
\label{eq20}
\end{equation}

Now we substitute (\ref{eq20}) and the respective known expression for the
TE contribution\cite{4} into (\ref{eq16}). Calculating the sum with the help
of the Abel-Plana formula, we obtain in perfect analogy to Ref.~\refcite{4}
\begin{eqnarray}
{\cal F}^{\rm mod}(a,T)&=&{\cal F}_{gp}(a,T)-\frac{k_BT}{16\pi a^2}
\int_{0}^{\infty}\!\!\!y\,dy\,\ln\left[1-r_{{\rm TE},gp}^2(0,y)\,
{\rm e}^{-y}\right]
\nonumber \\
&+&{\cal F}^{(\gamma)}(a,T)+
\beta_a{\cal F}^{(\beta)}(a,T),
\label{eq21}
\end{eqnarray}
\noindent
where ${\cal F}^{(\gamma)}(a,T)$ is determined by Eq.~(17)
in Ref.~\refcite{4}.
It goes to zero together with its derivative with respect to temperature,
when $T\to 0$. The quantity ${\cal F}^{(\beta)}(a,T)$ originates from
the second contributions on the right-hand sides of (\ref{eq19}),
(\ref{eq20}). Using the Abel-Plana formula,\cite{3}
it can be easily seen that
${\cal F}^{(\beta)}(a,T)=E^{(\beta)}(a)+O(T^3/T_{\rm eff}^3)$ at low $T$.
The Casimir free energy ${\cal F}_{gp}(a,T)$ is defined by substituting the
dielectric permittivity of the generalized plasma-like model
[Eq.~(\ref{eq5}) with $\gamma=0$]
into the Lifshitz formula. It was found in Refs.~\refcite{11,12}, and the
respective thermal
correction was shown to be of order $(T/T_{\rm eff})^3$ when $T\to 0$.
The TE reflection coefficient calculated using the generalized plasma-like
model at zero frequency entering (\ref{eq21})
is given by
\begin{equation}
r_{{\rm TE},gp}(0,y)=\frac{cy-\sqrt{4a^2\omega_p^2+c^2y^2}}{cy
+\sqrt{4a^2\omega_p^2+c^2y^2}}.
\label{eq22}
\end{equation}

Finally, calculating the Casimir entropy with inclusion of the screening
effects,
\begin{equation}
S^{\rm mod}(a,T)=-\frac{\partial{\cal F}^{\rm mod}(a,T)}{\partial T},
\label{eq23}
\end{equation}
\noindent
using Eq.~(\ref{eq21}) in the limit $T\to 0$, one obtains
\begin{equation}
S^{\rm mod}(a,0)=\frac{k_B}{16\pi a^2}
\int_{0}^{\infty}\!\!\!y\,dy\,\ln\left[1-r_{{\rm TE},gp}^2(0,y)\,
{\rm e}^{-y}\right]<0.
\label{eq24}
\end{equation}
\noindent
Thus, the Nernst heat theorem is violated and the theoretical approach
leading to the modified reflection coefficients is thermodynamicly
inconsistent. Note that this result is obtained for metals with perfect
crystal lattices. In the presence of impurities the Casimir entropy
abruptly jumps to zero\cite{20} at $T<10^{-3}\,$K. This, however, does
not solve the problem of consistency with quantum statistical physics
as discussed in Sec.~2.

Next we consider the low-temperature behavior of the Casimir free energy
for dielectric materials which includes screening effects. This is
also relevant for semiconductors with concentration of charge carriers below
the critical value. For these materials $n$ is relatively small and
$\kappa=\kappa_{\rm DH}$, as defined in Eq.~(\ref{eq13}),
should be used. The derivation
of the low-temperature behavior can be performed as in the case
of two dielectric semispaces with the inclusion of dc conductivity.\cite{5}
For dielectric materials the small parameter $\beta_l$ is given by
\begin{equation}
\beta_l=\frac{4\pi\sigma(0)}{\xi_l}\quad(l\geq 1),
\qquad
\sigma(0)\sim\exp\left(-\frac{C}{k_BT}\right),
\label{eq25}
\end{equation}
\noindent
where $C$ is some constant having a different meaning for different classes
of dielectrics. The parameter $\beta_l$ goes to zero when the temperature
vanishes. Then we expand the modified reflection coefficients
$r_{\rm TM,TE}^{\rm mod}({\rm i}\zeta_l,y)$ with $l\geq 1$ in powers
of the small parameter $\beta_l$,
\begin{eqnarray}
&&
{r}_{\rm TM}^{\rm mod}({\rm i}\zeta_l,y)={r}_{\rm TM}({\rm i}\zeta_l,y)+
\beta_l\frac{\varepsilon_l y[2y^2+(\varepsilon_l-2)\zeta_l^2]}{\sqrt{y^2+
(\varepsilon_l-1)\zeta_l^2}[\varepsilon_l y+\sqrt{y^2+
(\varepsilon_l-1)\zeta_l^2}]^2} +O(\beta_l^2),
\nonumber \\[-2mm]
&& \label{eq26}\\[-2mm]
&&
{r}_{\rm TE}^{\rm mod}({\rm i}\zeta_l,y)={r}_{\rm TE}({\rm i}\zeta_l,y)+
\beta_l\frac{y[y-\sqrt{y^2+(\varepsilon_l-1)\zeta_l^2}]}{\sqrt{y^2+
(\varepsilon_l-1)\zeta_l^2}[ y+\sqrt{y^2+
(\varepsilon_l-1)\zeta_l^2}]} +O(\beta_l^2).
\nonumber
\end{eqnarray}
\noindent
Here, the reflection coefficients $r_{\rm TM,TE}$ are defined by
Eq.~(\ref{eq2}) with the dielectric permittivity of core electrons (\ref{eq4}).
The Casimir free energy ${\cal F}(a,T)$ calculated with the coefficients
$r_{\rm TM,TE}$ vanishes with temperature as $\sim T^3$ (see Ref.~\refcite{5}).

Now we substitute (\ref{eq26}) in Eq.~(\ref{eq16}) and arrive at the following
expression for the Casimir free energy taking the screening effects into
account:
\begin{equation}
{\cal F}^{\rm mod}(a,T)={\cal F}(a,T)+\frac{k_BT}{16\pi a^2}\left\{
\int_{0}^{\infty}\!\!\!y\,dy\ln\left[1-
{r}_{\rm TM}^{\rm mod}{\vphantom{r^{\rm d}}}^2(0,y){\rm e}^{-y}\right]
+{\rm Li}_3(r_0^2)+Q(T)
\right\},
\label{eq27}
\end{equation}
\noindent
where ${\rm Li}_n(z)$ is the polylogarithm function,
$Q(T)$ vanishes exponentially\cite{5} when $T\to 0$,
$r_0\equiv(\varepsilon_0-1)/(\varepsilon_0+1)$ and
\begin{equation}
{r}_{\rm TM}^{\rm mod}(0,y)=
\frac{\varepsilon_0\sqrt{y^2+(2a\kappa_{\rm DH})^2}-
y}{\varepsilon_0\sqrt{y^2+(2a\kappa_{\rm DH})^2}+y}
\label{eq28}
\end{equation}
\noindent
is the reflection coefficient (\ref{eq9}) expressed in terms of the
dimensionless variables.
Calculating the negative derivative of both sides of (\ref{eq27})
with respect to $T$,
we obtain the asymptotic behavior of the Casimir entropy at low
temperature
\begin{eqnarray}
&&
{S}^{\rm mod}(a,T)=S(a,T)-\frac{k_B}{16\pi a^2}\left\{
\vphantom{\int\limits_{0}^{0}}
\int_{0}^{\infty}\!\!\!y\,dy\ln\left[1-
{r}_{\rm TM}^{\rm mod}{\vphantom{r^{\rm d}}}^2(0,y){\rm e}^{-y}\right]
+{\rm Li}_3(r_0^2)\right.
\nonumber \\
&&~
-8a^2\varepsilon_0T\frac{\partial\kappa^2}{\partial T}
\int_{0}^{\infty}\!\!\!dy
\frac{y^2 {r}_{\rm TM}^{\rm mod}{\vphantom{r^{\rm d}}}^2(0,y)}{{\rm e}^{y}-
{r}_{\rm TM}^{\rm mod}{\vphantom{r^{\rm d}}}^2(0,y)
}\,\frac{1}{\sqrt{y^2+(2a\kappa_{\rm DH})^2}[\varepsilon_0
\sqrt{y^2+(2a\kappa_{\rm DH})^2}+y]^2}
\nonumber \\
&&~\left.
+Q(T)+TQ'(T)
\vphantom{\int\limits_{\infty}^{\infty}}\right\},
\label{eq29}
\end{eqnarray}
\noindent
where $S(a,T)$ is defined using ${\cal F}(a,T)$ and, thus, vanishes
when $T\to 0$. It is easily seen that the last three
terms in curly brackets  on the right-hand side of this
equation go to zero when $T$ goes to zero for any dielectric material.

The behavior of the first two terms in the curly brackets on the right-hand
side of (\ref{eq29}) when $T$ goes to zero is more involved.
If $n(T)$ exponentially decays to zero with vanishing temperature
(as is true for pure insulators and intrinsic semiconductors), then
according to (\ref{eq13}) so does $\kappa_{\rm DH}$. As a result,
${r}_{\rm TM}^{\rm mod}(0,y)\to r_0$ and the
first two terms in the curly brackets cancel. Then the Casimir entropy
${S}^{\rm mod}(a,T)$ goes to zero when $T$ vanishes following the same law
as $S(a,T)$, i.e., as $T^2$. This means that for insulators and intrinsic
semiconductors the formalism under consideration is in agreement with the
Nernst heat theorem.

However, there is a wide class of dielectric materials (such as doped
semiconductors with dopant concentration below critical,
dielectric like semimetals, certain  amorphous
semiconductors, and solids with
ionic conductivity) for which $n$ does not go to zero when $T$ goes to zero.
 Although $\sigma(0)$
goes to zero exponentially fast for all dielectrics when $T$ goes to
zero, for most of them this happens due to the vanishing mobility
[see Eq.~(\ref{eq6})].
For instance,  the conductivity of SiO${}_2$ discussed in
Ref.~\refcite{14} is ionic in nature and is
determined by the concentration of impuritues. For all such materials,
in accordance with Eq.~(\ref{eq13}), $\kappa_{\rm DH}\to\infty$ when $T\to 0$.
As a result, ${r}_{\rm TM}^{\rm mod}(0,y)\to 1$ when $T$ goes to zero
in accordance
with (\ref{eq28}). In this case we obtain from (\ref{eq29})
\begin{equation}
{S}^{\rm mod}(a,0)=\frac{k_B}{16\pi a^2}\left[\zeta(3)-
{\rm Li}_3(r_0^2)\right]>0,
\label{eq31}
\end{equation}
\noindent
i.e., the Casimir entropy is positive and depends on separation distance
in violation of the Nernst heat theorem. This means that for a wide
class of dielectric materials the proposed approach taking the screening
effects into account is thermodynamically
inconsistent\cite{36}\cdash\cite{39}
in the same way as
the standard Lifshitz theory with the dc conductivity included.

We emphasize that the existence of dielectric materials for which $n$
does not go to zero but $\mu$ does go to zero when $T$ vanishes
demonstrates that the reflection coefficient (\ref{eq17}) at $\xi=0$ is
ambiguous. In reality, for such materials
${r}_{\rm TM}^{\rm mod}(0,k_{\bot})\to 1$ when $T$ and $\mu$ simultaneously
vanish. This is because $\kappa_{\rm DH}\to\infty$ when $T\to 0$
in disagreement with
physical intuition that there should be no screening at zero mobility.
This ambiguity is connected with the break of continuity of
the reflection coefficient ${r}_{\rm TM}^{\rm mod}({\rm i}\xi,k_{\bot})$
at the point $\xi=0$, $T=0$.
If one takes the limit $T\to 0$ first, keeping
$\xi={\rm const}\neq 0$, the standard Fresnel reflection coefficients
${r}_{\rm TM}$ from Eq.~(\ref{eq2})  with no screening are reproduced.
This property is preserved in the subsequent limiting
transition $\xi\to 0$. Thus, as was already noted above,
the violation of the Nernst heat theorem is caused by the break
of continuity of the reflection coefficients at the point (0,0) of the
$(\xi,T)$-plane.\cite{33}

Recently it was claimed\cite{16} that the nonlocal approach
leading to the reflection coefficient (\ref{eq9}) with
$\kappa=\kappa_{\rm DH}$ satisfies the Nernst
theorem, specifically, for solids with ionic conductivity which
is the conductivity of
activation type.
To prove this, Ref.~\refcite{16} arbitrarily
 separates the
thermal dependence of $\sigma(0)$ in Eqs~(\ref{eq6}), (\ref{eq25})
from the mobility $\mu$ and attributes it to the ``effective
density of charges, which are able to move".  This transfer of the
temperature dependence from $\mu$ to $n$ is incorrect\cite{36} because
the commonly used density of charge carriers $n$ producing the effect
of screening in
ionic conductors is an independently measured quantity,
which does not vanish with $T$.
Independent measurements of all three quantities, conductivity,
charge carrier concentration and mobility, demonstrate that
``mobility has the dominating influence upon the
conductivity-temperature dependence".\cite{37}
Nevertheless, Ref.~\refcite{40} expresses doubts concerning
the existence of dielectric materials whose density of charge
carriers does not go to zero when $T$ vanishes, while the
conductivity goes to zero due to vanishing mobility.
It is common knowledge, however, that for semimetals of the
dielectric type the Fermi energy is at a band where the
density of states is not equal to zero. The number of charge
carriers (electrons) near the Fermi surface in such dielectric
materials is fixed because it is determined by the structure
of the crystal lattice. Thus, the density of charge carriers
is nonzero at any $T$ including $T=0$.\cite{41}
The same holds for certain of the amorphous
semiconductors\cite{42} and for doped semiconductors with
dopant concentration below critical.

Recent Ref.~\refcite{34a} claims that it explicitly shows
the satisfaction of the Nernst theorem in the Lifshitz
theory with the modified reflection coefficients
$r_{\rm TM,TE}^{\rm mod}$ introduced in Ref.~\refcite{15}.
According to Ref.~\refcite{34a}, the Nernst theorem is
satisfied ``in systems with low density of carriers
(intrinsic semiconductors, dielectrics, etc)".
However, the presented proof uses an assumption that
``the carrier density vanishes as $T\to 0$".\cite{34a}
Thus, wide classes of dielectric materials for which
the density of charge carriers does not vanish when
the temperature vanishes (doped semiconductors with dopant
concentration below critical, semimetals of dielectric type,
some amorphous semiconductors and solids with ionic
conductivity) are simply excluded from consideration.
As a result, the proof of Ref.~\refcite{34a} is in fact
applicable to only insulators and intrinsic semiconductors,
where the Fermi energy at $T=0$ lies in a band gap.
For these materials the density of states at the Fermi energy
is equal to zero. Thus, for insulators and intrinsic
semiconductors the Nernst theorem is satisfied, as was proved
above on the basis of Eq.~(\ref{eq29}). However, for the
majority of dielectric materials $n$ does not go to zero
with vanishing $T$ leading to the violation of the Nernst
theorem in the modification of the Lifshitz theory
proposed in Ref.~\refcite{15} and also in
Refs.~\refcite{14},\,\,\refcite{16}.

There is a remark in the literature\cite{18a} that the
violation of Nernst's theorem for dielectric materials
where $n$ does not go to zero with vanishing $T$ is
``a pure misunderstanding" because ``The materials under
discussion are amorphous glass-like {\it disordered}
bodies" for which ``Nernst's theorem {\it is not valid}."
This remark is erroneous in two aspects.
First, as discussed above, there are dielectric materials
with an ordered structure for which $n$ does not vanish
with $T$ (dielectric like semimetals, for instance).
Second, the violation of Nernst's theorem for disordered
bodies is irrelevant to the Casimir entropy discussed
here. As is correctly stated in Ref.~\refcite{18a},
nonzero entropy of glass plates at $T=0$ is connected
with the fact that they are simply not at an
equilibrium state at low $T$. This entropy does not
depend on separation between the plates and, thus,
is not in contradiction with the Nernst theorem.
By contrast, the Casimir entropy (\ref{eq31})
is nonzero and
depends on separation distance (i.e., on the volume
of the system) for both ordered and disordered
materials of plates (provided $n$ does not vanish with
$T$) which is in contradiction with the Nernst theorem.

Thus, the substitution of the modified reflection coefficients into
the Lifshitz formula leads to contradictions with thermodynamics.
The physical reason for this is that the Lifshitz theory describes
a system in thermal equilibrium whereas the modified reflection
coefficients were obtained for a system that includes both
drift and diffusion
currents which are irreversible processes out of thermal equilibrium.
As was emphasized at the beginning of this section, the Boltzmann transport
equation describes irreversible processes which occur, for instance,
in an external electric field. It is a far reaching extrapolation to apply
this equation to the fluctuating electromagnetic field, as is done in
Ref.~\refcite{15}. In an external electric field the system goes out of
thermal equilibrium and the fluctuation-dissipation theorem is violated.
This cannot happen and does not happen in the presence of
fluctuating electromagnetic fields. One can conclude that the thermodynamic
puzzles discussed above are artifacts of the application of the Lifshitz
theory and Boltzmann transport equation outside of their
range of applicability.

\section{Experimental Tests for the Influence of Screening
Effects on the Casimir Force}

The theoretical predictions made by the approaches including screening
effects in the Lifshitz theory can be compared with the recent experimental
results from the measurement of the Casimir force between metal-metal
and metal-semiconductor test bodies. In fact all three theoretical
approaches\cite{14}\cdash\cite{16} discussed above lead to almost
coincident predictions. This is because the contributions from the zero
frequency term in all these approaches are exactly the same and the
contributions from all nonzero Matsubara frequencies are approximately
equal.

We begin with the most precise experiment on the indirect measurement
of the Casimir pressure between two gold plates by means of a
micromechanical torsional oscillator.\cite{7,8} In this experiment,
the configuration of a sphere above a plate was used in the dynamic regime
to determine the equivalent Casimir pressure in the configuration of
two plates using the proximity force approximation. (This experiment
was already mentioned in Sec.~2 in the comparison between the Drude model
and the generalized plasma-like model in combination with the standard
Lifshitz theory.) Note that Fig.~1(a) can be also used to compare the
experimental data shown as crosses with the predictions of the theoretical
approaches taking screening effects into account.\cite{14}\cdash\cite{16}
The point is that at separations larger than 300\,nm the approaches including
the screening effects with $\kappa=\kappa_{\rm TF}$ lead to the same
computational results for the Casimir pressure as the Drude model approach
shown as the dark-gray band (the latter is plotted in the region from
500 to 600\,nm). As is seen in Fig.~1(a), the approaches taking into account
screening effects are experimentally excluded at a 95\% confidence level.
\begin{figure}[pt]
\vspace*{-0.8cm}
\centerline{\hspace*{-2cm}\psfig{file=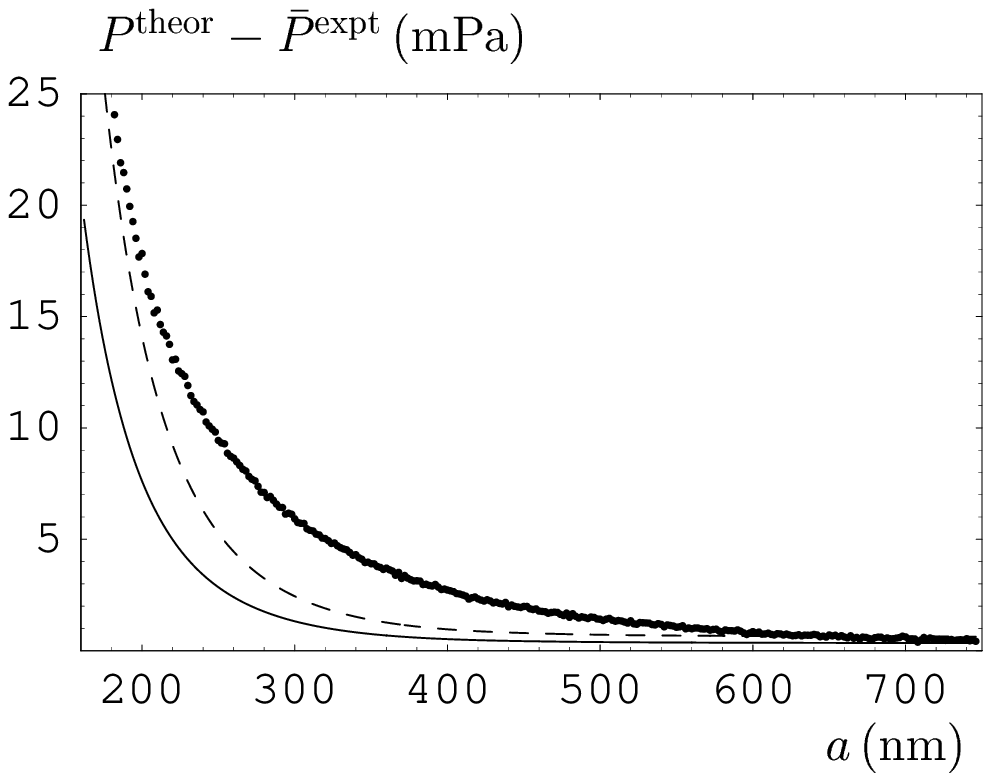,width=14cm}}
\vspace*{-14.cm}
\caption{ The differences of the
theoretical and the  mean experimental
Casimir pressures between the two Au plates
versus separation are shown as dots.
The theoretical results are calculated using the modified
Lifshitz theory with the inclusion of screening effects.
The solid and dashed lines indicate the boundaries
of  95\% and  99.9\% confidence intervals, respectively.}
\end{figure}
In Fig.~3 we use another method of comparison of these theoretical approaches
with data over the entire measurement range from 162 to 750\,nm. Here, the
differences between the theoretical and
the mean experimental Casimir pressures
are shown as dots. The solid and dashed lines indicate the boundaries
of  95\% and  99.9\% confidence intervals, respectively.
As is seen in the figure, the dots are outside the 95\% confidence interval
over the entire measurement range. At separations from 162 to 640\,nm
the theoretical approaches including the screening effects are
experimentally excluded at the 99.9\% confidence level.
Recently it was claimed\cite{40} that the electrostatic
calibrations in the experiments
of Refs.~\refcite{7},\,\,\refcite{8} do not take into account
``important systematic effects". This claim remains
unjustified until some specific objection to the calibration
procedures of Refs.~\refcite{7},\,\,\refcite{8} described in more
detail in Refs.~\refcite{43},\,\,\refcite{44} is presented.
It was claimed also\cite{45} that the comparison of experiment
with theory, like in Figs.~1,\,\,3, is irrelevant because the
optical properties of the Au films used were not measured but
taken from tables. However, computations performed in
Ref.~\refcite{46} demonstrated that the use of any alternative
set of optical data only increases disagreement between the
Drude model approach or approaches of
Refs.~\refcite{14}--\refcite{16} and the experimental data.

Next we compare the theoretical predictions with
the inclusion of the screening
effects with the measurement data of the experiment on the modulation
of the Casimir force with light.\cite{26,27}
(This experiment was also mentioned in Sec.~2.) Here, the difference
of the Casimir forces $F_{\rm diff}$ between an Au coated sphere and a Si
plate was measured in the presence, and in the absence, of laser light on
the plate. In the absence of light, the concentration of charge carriers
in Si was much below the critical value, i.e., Si was in a dielectric state.
In the presence of light, the concentration of charge carriers
in Si was above the critical value, i.e., Si was in a metallic state
(all values of respective parameters are listed in Ref.~\refcite{27}).

The theoretical predictions with the inclusion of screening effects are
numerically almost the same, as in the standard Lifshitz theory, for the
case when the Si plate is in the bright phase, but differ measurably when
the Si plate is in the dark phase in comparison with the calculation
where the conductivity of Si in the dark phase is simply neglected
(see Sec.~2). It is not possible, however, to conclusively compare
experiment with theory taking the screening effects into account
at a 95\% confidence level. Below we compare the experimental data
whose total experimental errors are determined at a 70\% confidence level
with the theoretical bands computed with inclusion of screening effects.
Note that in the dark phase $\kappa=\kappa_{\rm DH}$ was used in the
computations. In the bright phase the computational results for the
Casimir force are almost the same for both $\kappa=\kappa_{\rm DH}$ and
$\kappa=\kappa_{\rm TF}$. The width of the bands is found at the same
70\% confidence level as the experimental errors.
This width is mostly determined by the error in the charge
carrier densities used in the computations.
In Fig.~4(a,b) the experimental difference Casimir forces,
$F_{\rm diff}^{\rm expt}$ (the force in the bright phase minus the force
in the dark phase) are shown as crosses for different measurements with
absorbed power $P_w=9.3\,$mW and $8.5\,$mW, respectively. The theoretical
bands for $F_{\rm diff}^{\rm theor}$ computed with
the inclusion of the screening
effects lie in between the dashed lines.
\begin{figure}[t]
\vspace*{-4.5cm}
\centerline{\hspace*{-15mm}\psfig{file=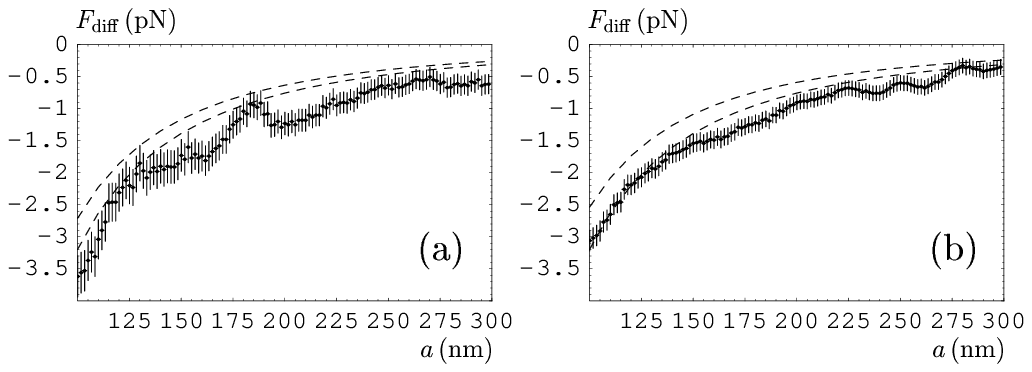,width=24cm}}
\vspace*{-25.3cm}
\caption{(a) The measured force differences are shown as crosses
versus separation $a$ for the absorbed power (a) $P_w=9.3\,$mW
and (b) $P_w=8.5\,$mW. The theoretical results calculated with
inclusion of screening effects lie between the  dashed lines.
}
\end{figure}
As is seen in Fig.~4(a,b), the theoretical approach taking into account
the screening effects is excluded by the data at almost all separation
distances within the measurement range from 100 to 300\,nm.

In Fig.~5 the experimental data from one more repetition of the same
experiment with a lower absorbed power ($P_w=4.7\,$mW) are compared with
the theoretical approaches taking the screening effects into account.
In Fig.~5(a) the experimental data are shown as crosses. The theoretical
results computed with the inclusion of the screening effects belong to the
band in between the two dashed lines. As is seen in Fig.~5(a), below 200\,nm
and above 275\,nm the theory is inconsistent with data. In Fig.~5(b) the same
data are compared with the same theoretical approach using another method
of comparison.
\begin{figure}[pb]
\vspace*{-4.7cm}
\centerline{\hspace*{-15mm}\psfig{file=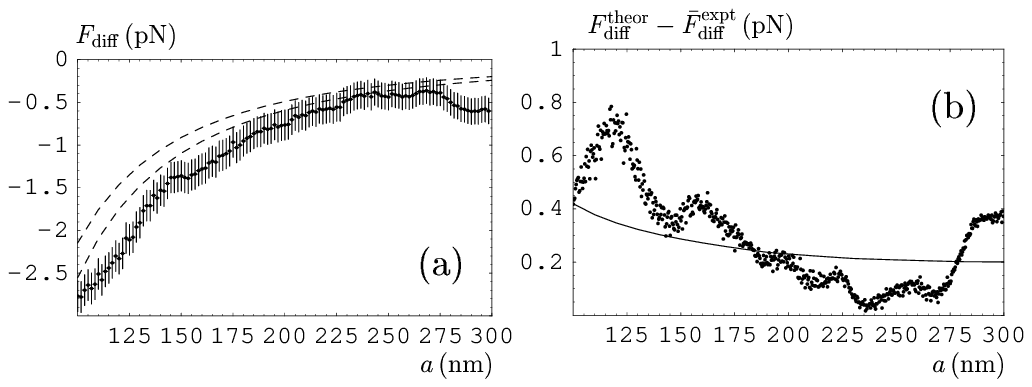,width=24cm}}
\vspace*{-25.3cm}
\caption{(a) The measured force differences are shown as crosses
versus separation $a$ for the absorbed power $P_w=4.7\,$mW.
The theoretical results calculated with
inclusion of screening effects lie between the  dashed lines.
(b) Theoretical minus mean experimental differences of the
Casimir force for the Lifshitz theory with
inclusion of screening effects.
The solid line indicates 70\% confidence intervals.
}
\end{figure}
Here, theoretical  minus experimental differences of the Casimir forces
are plotted as dots. The solid line shows the upper boundary of the 70\%
confidence intervals taking into account all experimental and theoretical
errors. It is seen that below 185\,nm and above 280\,nm all dots are
outside the confidence intervals. As a result, Figs.~4(a,b) and 5(a,b)
conclusively indicate that the theoretical approaches taking into account
screening effects are excluded by the data of the optical modulation
experiment\cite{26,27} at a 70\% confidence level.

In contrast to this conclusion, it was claimed\cite{16} that
``the experiment can hardly distinguish between nonlocal theory and
local theory with zero conductivity of Si.'' (We recall that according
to Sec.~2, Fig.~2, the local theory with dc conductivity of Si
neglected in the dark phase is in perfect agreement with the data.)
The comparison between the theoretical and experimental results in
Ref.~\refcite{16} is, however, irregular.\cite{36}
In Fig.~1(a) of Ref.~\refcite{16} the experimental data
[taken from Fig.~1(a) of Ref.~\refcite{28} with no indication of the
source] are shown with errors determined at a 70\%
confidence level. In the same figure the theoretical band for the nonlocal
approach including the screening effects is obtained from the uncertainty
in  $n$,
$\Delta n=0.4\times 10^{19}\,\mbox{cm}^{-3}$, determined\cite{27}
at a 95\% confidence level.
The reader of Ref.~\refcite{16} is not informed about the
confidence levels used.
This comparison of experiment with theory
is thus confusing. In our Fig.~4(a)
the same data are compared with the predictions
of the nonlocal approach (the band between the dashed
 lines), where the band width is determined
 at the same 70\% confidence level, i.e., with
$\Delta n=0.3\times 10^{19}\,\mbox{cm}^{-3}$,
as the errors of the data.
It can be clearly seen that  the nonlocal
approach is excluded by the data at a 70\% confidence level.

Regarding Fig.~1(b) in Ref.~\refcite{16}, it shows
both the data and the theoretical bands  at
a 95\% confidence level and restricts the region of separations only
from 100 to 150\,nm. However,\cite{28} these  data
cannot be conclusively compared with the nonlocal approach at such a
high confidence.
The correct comparison between the data and the nonlocal approach for this
measurement set
of the optical modulation experiment over the entire separation
range is presented in our Fig.~4(b). It can be clearly seen that data are
inconsistent with the nonlocal approach including the screening effects over
a wide range of separations.

\section{Conclusions and Discussion}

In the above, we have reviewed problems with the Lifshitz theory
when the drift current of conduction electrons in metals, or the dc
conductivity at $T\neq 0$ in dielectrics, are included
in the model of the dielectric response. It was shown that these
phenomena violate thermal equilibrium which is the basic applicability
condition of the Lifshitz theory. Because of this, the inclusion of the
drift current into the Lifshitz formula results in thermodynamic
and experimental inconsistencies.

The attempts to avoid these problems introduce modified reflection
coefficients taking into account screening effects and diffusion
currents.\cite{14}\cdash\cite{16} We show (see
also Refs.~\refcite{36}--\refcite{39}) 
that the Lifshitz formula combined with the
modified reflection coefficients leads to nonzero Casimir entropy at 
$T=0$ depending on the separation distance between the plates.
Thus, the proposed approaches violate the third law of thermodynamics.
These approaches are also demonstrated to be
inconsistent with the measurement data of two experiments.
For metal-metal test bodies they are excluded
experimentally at a 99.9\% confidence level. In the configuration of
metal-semiconductor test bodies
the exclusion is confirmed at a 70\% confidence level.

The reason for the failure of the theoretical approaches,  including
the screening effects and diffusion currents into the Lifshitz theory,
is that these processes violate
thermal equilibrium which is the basic applicability
condition of the Lifshitz theory. Both the drift and diffusion
currents are irreversible phenomena which take place out of thermal
equilibrium. By contrast, the dispersion forces are
physical phenomena of fluctuating nature which go on in thermal
equilibrium. One arrives at the conclusion that there is a deep
difference between external and fluctuating electromagnetic fields which
might not be sufficiently reflected in the mathematical formalism of
quantum theory. Future studies will show how important this conclusion is
for wider ranges of physical phenomena beyond the scope of dispersion
forces.

\section*{Acknowledgments}

R.S.D.~acknowledges NSF support through Grants No.~CCF-0508239
and PHY-0701636, and from the Nanoscale Imaging Center at IUPUI.
E.F. was supported in part by DOE under Grant No.~DE-76ER071428.
U.M., G.L.K. and V.M.M. were
 supported by the NSF Grant No.~PHY0653657
(computations of the Casimir force) and
DOE Grant No.~DE-FG02-04ER46131 (measurement of the Casimir
force differences).
G.L.K. and V.M.M. were also partially supported by
Deutsche Forschungsgemeinschaft, Grant No.~436\,RUS\,113/789/0--4.

\end{document}